# Remaining useful life prediction with uncertainty quantification: development of a highly accurate model for rotating machinery

Zhaoyi Xu, Yanjie Guo, Joseph Homer Saleh

*School of Aerospace Engineering, Georgia Institute of Technology, Atlanta, GA 30332, USA*

**Abstract:** Rotating machinery is essential to modern life, from power generation to transportation and a host of other industrial applications. Since such equipment generally operates under challenging working conditions, which can lead to untimely failures, accurate remaining useful life (RUL) prediction is essential for maintenance planning and to prevent catastrophic failures. In this work, we address current challenges in data-driven RUL prediction for rotating machinery. The challenges revolve around the accuracy and uncertainty quantification of the prediction, and the non-stationarity of the system degradation and RUL estimation given sensor data. We devise a novel architecture and RUL prediction model with uncertainty quantification, termed VisPro, which integrates time-frequency analysis, deep learning image recognition, and nonstationary Gaussian process regression. We analyze and benchmark the results obtained with our model against those of other advanced data-driven RUL prediction models for rotating machinery using the PHM12 bearing vibration dataset. The computational experiments show that (1) the VisPro predictions are highly accurate and provide significant improvements over existing prediction models (three times more accurate than the second-best model), and (2) the RUL uncertainty bounds are valid and informative. We identify and discuss the architectural and modeling choices made that explain this excellent predictive performance of VisPro.

# 1. Introduction

The concept of remaining useful life (RUL) refers to the expected remaining lifespan of a component or equipment. Its prediction is used to inform maintenance decisions and to minimize the risk of catastrophic failures during operation [1]. Accurate RUL prediction is critical for prognostic and health management (PHM) and maintenance planning. Preventive maintenance actions are scheduled based on RUL predictions and equipment's health condition to prevent unexpected failures and reduce total maintenance costs [2-4].

This work develops and validates a novel computational architecture and prognostic model for highly accurate RUL prediction of rotating machinery. Rotating machinery is essential to modern life. For example, power generation, transportation, and a host of other industrial applications are fundamentally dependent on rotating parts [5]. Rotating machinery generally operates under challenging working conditions, which can lead to untimely failures. As a result, failure prediction and timely maintenance are critical for the reliability of such equipment [6]. With the significant advances in sensor technologies and data acquisition in the last decade, data-driven methods for RUL prediction have become increasingly popular [7-13], and different machine learning RUL models have been proposed and applied to rotating machinery [14-22]. For example, Zhang et al [20] developed deep learning (DL) and health indicator based RUL prediction method and reported a high level of prediction accuracy. Wang et. al [19] developed a hybrid prognostic model with relevant vector machine (RVM) and health indicator for the RUL prediction of rolling element bearings, and they reported better prediction accuracy compared with traditional methods.

Despite the recent progress in this area, several challenges remain in data-driven RUL prediction for rotating machinery. The first challenge is related to the time-dependence of the system degradation and RUL prediction given vibration data. Current prognostic models make RUL predictions based on the present sensor measurements without accounting for the history of operation of the system. Sensor measurements, however, rarely if ever capture the entirety of the state vector of a system, and as a result, two similar measurements from the suite of sensors, for example, arrived at through different trajectories or degradation paths, do not necessarily reflect the same underlying health condition of the system. Making RUL predictions that do not account for time-dependence in sensor measurements can be inaccurate and misinform maintenance decisions, which in turn can fail to prevent run-to-failures. The second challenge is how to tease out more hidden information in the vibration raw data to improve the model prediction accuracy. This challenge includes two aspects. The first aspect relates to the form of the input data and the preprocessing of the original signal. Traditional models use the variance and magnitude of the machine vibration signal to construct a health index (HI) and perform the RUL prediction. However, the variance is not informative enough for accurate RUL prediction, and it is blind to information in the frequency domain. To address this drawback, some recent works [19, 20] combined information in the time and frequency domain as the input of their prognostic model. For example, Zhang et al [20] combined information in the time and frequency domains to form a $100 \times 100$ dimension input at each time step as the input of their prognostic model. The second aspect of this challenge relates to the model design and training to extract more useful information for accurate RUL prediction from the input data. Different types of models, such as CNN [18] and RVR [19], are used in the rotating machinery prognostic. Problems of the model design and training, such as model overfitting and underfitting [23-25], can degrade the accuracy of the prediction. These two aspects of this challenge are co-dependent. For example, the design and selection of a prognostic model should depend on the form of the input raw data to avoid overfitting and underfitting [23-25]. A more complex model is designed to extract more hidden information from the high dimensional and more informative input data.

This is a vigorous research area and several approaches have been proposed to tackle these challenges. We provide next a cursory overview of this literature. There is a broader context within which research on RUL prediction is situated. It is related to advances in Machine Learning (ML) in general, in Deep Learning (DL) and Gaussian Processes in particular for reliability and safety applications. We recently provided a review of this broad analytical landscape, and we include here a short excerpt for the convenience of the reader. More details can be found in [26]. Among the various ML technologies, deep learning (DL) is a key enabler of applications and analysis. It uses deep multi-layered neural architecture, such as fully connected and convolutional layers, to fulfill complex regression, classification, and some unsupervised functions. DL can improve model accuracy and tease out more latent information from the raw data compared with shallow ML methods. DL is increasingly used in the analysis of system degradation and RUL prediction in large part because of its superior ability to handle high-dimensional data. For example, Peng et. al. [27] developed a deep learning model with fully connected and convolutional layers to predict the RUL of the ball bearing machines and reported higher accuracy compared with traditional models and shallow machine learning models. Fink *et* al. [28] devised multilayer feedforward DL models for RUL and reliability prediction of complex systems and reported high-level accuracy in their validation tests using railway data. Another important tool that is used for RUL prediction is Gaussian process regression (GPR). GPR is a nonparametric, statistical, and Bayesian approach to model the evolution of the time streaming data with an uncertainty quantification over time. For example, Lederer et al. [29] developed a real-time GPR model for online RUL prediction and reported better accuracy compared with other models. A more extensive review of different ML approaches for RUL prediction can be found in [26]. While highly promising, these approaches do not fully address the two challenges noted previously, and consequently large inaccuracies in RUL prediction remain.

The main contribution of the present work is the development and validation of a novel computational architecture for a highly accurate RUL prediction model with uncertainty quantification. Our method adopts the advantages of DL neural network and non-stationary GPR to jointly address current challenges in the RUL prediction of rotating machinery. Our RUL prediction is tested and benchmarked using PHM12 real-word experimental dataset developed by the PROGNOSTIA platform [30], and results show significantly higher accuracy compared with other best-in-class alternatives in recent literature.

The remainder of the article is organized as follows. Our RUL prediction model and its technical details are introduced in Section 2. The datasets and computational experiments used to test and benchmark the performance of our RUL prediction model are presented in Section 3. The computational results are discussed in Section 4. Section 5 concludes this work.

**2. Prognostic and remaining useful life prediction model**

In this section, we first provide an overview of our prognostic and RUL prediction model. We refer to the model with the acronym VisPro: the "Vis" prefix is used to reflect the fact that the model uses DL image classification technology with the help of Time-Frequency Analysis (TFA) to visualize and examine the characteristics of the vibration signal over time and in the frequency domain.

## 2.1. Overview of the VisPro model

We begin with a high-level overview of the RUL model and leave the details to the following subsections. The general architecture of the model is shown in Fig. 1.

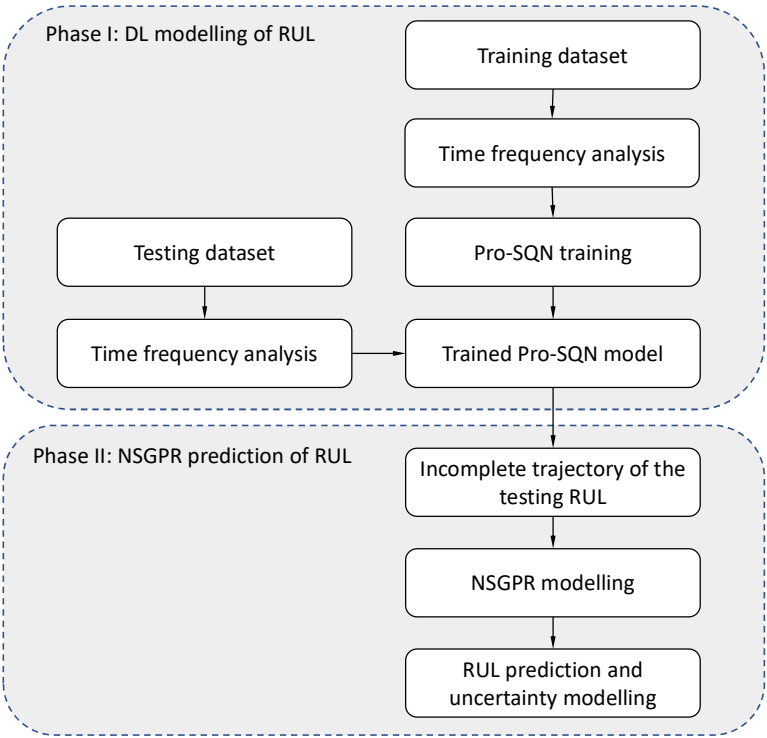

Figure 1. Overview of the VisPro model for RUL prediction

This VisPro model for RUL prediction consists of two phases, a first DL modeling phase, and a second nonstationary GPR prediction phase. In Phase I, we first conduct time-frequency analysis (TFA, more details in Section 2.2) on the input training data, which consists of a group of run-to-failure trajectories from the vibration data (i.e., up to the time to failure $t_f$). The run-to-failure enables the calculation of complete RUL trajectories up to $t_f$ in the training dataset (details in Section 3). The purpose of the TFA step is to expand the dimension of the original vibration signal to a 2D space of time-frequency domain to tease out more hidden information of the dataset. We then use the training data in the time-frequency domain to train the prognostic SqueezeNet (Pro-SQN) for the RUL estimation (details in Section 2.4). The trained Pro-SQN model at the end of Phase I predicts the RUL during system operation at every point in time $t < t_f$ when vibration data is available. In the testing dataset, which contains incomplete vibration signal trajectory (i.e., truncated before machine failure), the trained Pro-SQN updates its RUL prediction up to the time when the vibration data is truncated or censored at time $t_c$. For clarity, we index $RUL_t$ with the time at which the prediction is carried out. This concludes Phase I of the VisPro model.

As discussed in Ref. [26], there are two motivations for appending Phase II to this intermediate output. The first is a conceptual consideration and consists in the fact that the Phase I prediction is deterministic and lacks uncertainty estimation. The second is a practical consideration and consists in the fact that the Pro-SQN $RUL_t$ predictions exhibit oscillations of varying amplitudes over time, and this ultimately degrades the accuracy of the prediction. To achieve a more accurate and stochastic RUL prediction, we append Phase II and its NSGPR to our model. In Appendix A, we compare the RUL prediction with and without Phase II and discuss the advantage of its inclusion in VisPro.

The Pro-SQN model from Phase I predicts the $RUL_t$ trajectory from $t_0$ up to $t_c$ in the testing dataset, and this trajectory is fed to the NSGPR in Phase II. We use the NSGPR to model the evolution of these incomplete $RUL_t$ predictions in the testing dataset. By 'learning' from these predictions prior to the truncation time, the NSGPR captures the main trend in the RUL development over time. It is then used to provide more accurate and stochastic RUL predictions up to the time to failure along with uncertainty quantification around its prediction (details in Section 2.4). The reader interested in a result illustrating this statement (before examining the technical details of the VisPro) may jump ahead to Section 4 and glance at Fig. 9.

We designed this model to address the current challenges of data-driven RUL prediction discussed in the Introduction. First, to capture the time dependence of the system degradation and RUL prediction, we added the time variable to the input vector of the DL network (details in Section 3.3). Embedding the time variable in the input vector of the Pro-SQN model helps the network 'learn' the inherent time dependence of the degradation and RUL prediction. Second, to extract more useful information in the raw data and improve the prediction accuracy, we use the TFA preprocessing of the original vibration signal and design a Pro-SQN to estimate the RUL. We conduct TFA preprocessing to both the training and testing data.

In the next subsections, we discuss the technical details of the different pieces in our VisPro RUL prediction model.

### 2.2. Short-time Fourier transform time-frequency analysis

As shown in Fig. 1, the first step in VisPro is to use the time-frequency analysis (TFA) method to map the input time streaming vibration training and testing data into the time-frequency domain. We achieve this by leveraging a short-time Fourier transform (STFT) model. The STFT is a Fourier-related transform used to determine the frequency and phase content of local sections of a signal as it changes over time [31]. The basic idea of the STFT is to divides a time streaming signal into sequential time segments of equal length and then compute the Fourier transform separately on each segment and then plot the changing spectra as a function of time. To accomplish this, we introduce a time localization notion by using a sliding time window $w[·]$ going through the entire time streaming input signal $x[·]$. The Fourier analysis is performed to every windowed portion of the signal $x$ forming the frequency dimension. The sliding window forms the time dimension, and consequently, we formulate a 2D time-varying frequency analysis. The STFT result ($X$) representing the signal ($x$) in the time-frequency space is calculated by Eq. 1:

$$X(m,f) = \sum_{n} x[n]w[n-m]e^{-jfn} \tag{1}$$

where $x$ is the signal, $w$ is the window function, and $m, f$ are the variables for time and frequency dimension. Since we use fast Fourier transform (FFT) in each time segment, The parameters of the STFT result, $m, f$ are discrete variables [31]. The frequency resolution is equal to the inverse of the window function duration, and a detailed analysis of the tradeoff between window size and frequency resolution is discussed in Ref. [32]. We output the magnitude of $X$ as a one-channel TFA layer, which is a $64 \times 64$ matrix with 64 frequency bands generated at 64 different time windows [32]. This matrix $X$ is the input to the Pro-SQN DL RUL estimation, and it is discussed in the next subsection.

**2.3. Deep learning RUL estimation: prognostic-SqueezeNet**

In this subsection, we develop a deep network for the RUL estimation work: prognostic-SqueezeNet (Pro-SQN). Iandola et. al. [33] developed SqueezeNet and noted that it provided excellent classification accuracy, as good as that of the award-winning AlexNet [34] while requiring 50 times fewer parameters than AlexNet. In order to tailor SqueezeNet to our fault classification and further improve the accuracy of this general model in our application, we modified the original model and developed a dedicated prognostic-SqueezeNet (Pro-SQN). The structure of our Pro-SQN is shown in Fig. 2.

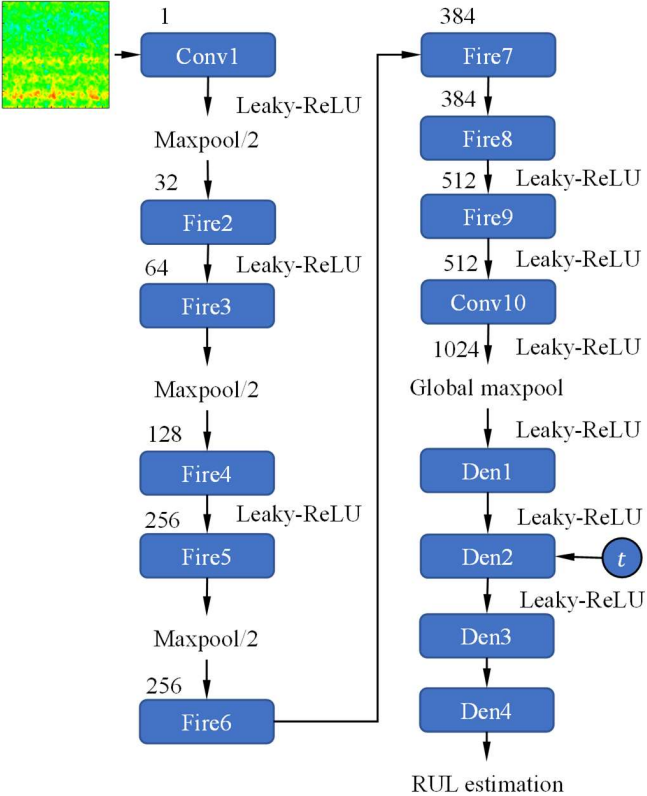

Figure 2. Structure of the Pro-SQN model for prognostic and RUL prediction

This Pro-SQN operates as follows. First, we input the TFA channel with a size of $64 \times 64 \times 1$ into the network. We then use convolutional layers, maxpoolings, and fire models (details in Appendix B) to map the input to a hidden vector of size 1024. We first use one dense layer (Den1) to transfer the output of the global maxpooling to four latent variables. We next embed the time variable to the input of "Den2" layer to capture the time dependency between system degradation and RUL estimation. Finally, we use three fully connected layers (Den2, Den3, and Den4) to convert the four hidden variables along with the time variable to the final RUL prediction at the corresponding time. We discuss the weight function of each layer and the overall memory size of this Pro-SQN in Appendix B. The weight size of this Pro-SQN model is 0.6Mb, and the overall processing memory per image is 4.7 Mb.

To adjust the original SqueezeNet to a prognostic model and improve its RUL prediction accuracy, we implemented the following modifications to the original model. First, we modified the classification problem with 1000 outputs in the original SqueezeNet to single outputs for the RUL estimation at the corresponding time. This is achieved by using fully connected dense layers. Second, since TFA signal evolution exhibits different phases (as shown in Fig. 7), to model the transition between different phases, we select Leaky-ReLU stepwise activation function. The use of the Leaky-ReLU can better model nonlinearities by avoiding the vanishing gradient problem in model training. Third, we embed the time variable as the input of this model by using it as one of the inputs in "Den2" layer. We use "Den2", "Den3", and "Den4" layers to integrate the information of four hidden variables representing sensor monitored machine condition and the time variable to achieve a more accurate RUL prediction.

### 2.4. Non-stationary Gaussian process regression RUL prediction

The reasons for appending Phase II and the NSGPR were discussed previously. The Gaussian process regression (GPR) is a Bayesian approach that can be used to model time streaming data. As shown in Fig. 1, the GPR surrogate model is used to model the evolution in time of the RUL prediction from Phase I. A brief introduction to GPR and our nonstationary kernel are provided next. For more details, the reader is referred to [35].

Consider $m$ realizations of a time streaming dataset in $m$ time steps. The input variables are denoted as $X_s = (x_1, x_2, \ldots, x_m)$, and each $x_i$ is a $d$ dimensional vector. The values of the output objective functions are denoted as $y_s = (y_1, y_2, \ldots, y_m)$, with each $y_i = Y(x_i)$, as the $m$ realizations of a stochastic function $Y(x)$, say RUL predictions. The GPR is formulated in Eq. 2, with $Y(x)$ decomposed into a deterministic mean approximation $f(x)$, and a stochastic part $Z(x)$. The stochastic part is a centered Gaussian process characterized by its covariance function. A quadratic function, as shown in Eq. 3, is used in this work to model the mean part $f(x)$, where $\boldsymbol{\beta}$ is the weight and bias matrix. The stochastic part covariance function ($Cov(Z(x_i), Z(x_j))$) is modeled by selected kernel functions, and we will discuss our choice of covariance functions shortly. With the mean regression, covariance kernel functions, and $Y(x)$ decomposition formulation, the GPR prediction of new input ($x_{new}$) is considered as a posterior Gaussian distribution, which is shown in Eq 8. The Gaussian distribution mean and variance values are calculated by Eq. 4, where $\sigma_e$ is the Gaussian modeled noise standard deviation.

$$Y(x) = f(x) + Z(x) \qquad (2)$$

$$\begin{cases} f(x) = H(x) \times \beta \\ \\ \quad \text{with} \\ \\ H(x) = [1, x, x^2] \\ x = [x_1, x_2, \ldots, x_d] \\ x^2 = [x_1^2, x_2^2, \ldots, x_d^2] \end{cases} \quad (3)$$

$$\mu_{new} = Cov(Z(X_s), Z(x_{new}))\left(Cov(Z(X_s), Z(X_s)) + \sigma_e^2 I\right)^{-1} \times$$

$$\left(Y(X_s) - f(X_s)\right) + f(x_{new})$$

$$\sigma_{new}^2 = Cov(Z(x_{new}), Z(x_{new})) Cov(Z(x_{new}), Z(X_s)) \times$$

$$\left(Cov(Z(X_s), Z(X_s)) + \sigma_e^2 I\right)^{-1} Cov(Z(X_s), Z(x_{new}))$$

(4)

Our nonstationary GPR covariance function consists of two parts: (1) a dot product kernel function and (2) a local length scale kernel. For the first part, we select a simple non-stationary kernel, dot product, to capture the nonstationarity of the response by using the actual position of the points of $x_i$ and $x_j$ ($x_i, x_j \in X$) instead of the separation of the points, $|x_i - x_j|$, of typical stationary kernels, such as squared exponential kernel [36]. The dot product kernel is shown in Eq. 5.

$$Cov_2\left(Z(x_i), Z(x_j)\right) = \sigma_0^2 + x_i x_j \quad (5)$$

Regarding the second part of a local scale length kernel, typically, traditional stationary kernels, such as squared exponential kernel [37], assume a constant length-scale across the entire input space and hence the covariance only depends on the distance between two input locations, say $|x_i - x_j|$. Intuitively, length-scale represents the realm of correlation between different locations. For example, if the output variable decreases sharply, farther points should not be affected significantly, and it requires a short length-scale locally. To account for the nonstationary properties of the output, an input dependent (local) length-scale is used [38]. This local length scale kernel is used successfully for 3-D digital terrain modeling [39] and environmental monitoring [40]. In this work, since the RUL and machine failure prediction in time is one-dimensional problem, the local length scale kernel is shown as Eq. 6 [38].

$$Cov_2\left(Z(x_i), Z(x_j)\right) = \sigma_f^2 (l_i^2)^{0.25} (l_j^2)^{0.25} (0.5l_i^2 + 0.5l_j^2)^{-0.5} \times \exp\left[-\frac{(x_i - x_j)^2}{0.5l_i^2 + 0.5l_j^2}\right] \quad (6)$$

Length-scale hyperparameter $l_i$ and $l_j$ are evaluated for every alternative location $x_i$ and $x_j$ for the input locally. The length scale is modelled by a second-level GP. In this work, we use a GP with a squared exponential kernel as the second-level GP to model the length scale. This local length scale kernel can significantly improve the prediction accuracy of the RUL in our experiment. The effect and improvement of using this local length scale kernel are discussed in Appendix C. For more details regarding this local length scale kernel, the reader is referred to [38].

We add this local length scale kernel $Cov_2$ to the dot product kernel $Cov_1$ for the overall covariance function in the NSGPR as Eq. 7.

$$Cov\left(Z(x_i), Z(x_j)\right) = Cov_1\left(Z(x_i), Z(x_j)\right) + Cov_2\left(Z(x_i), Z(x_j)\right) \quad (7)$$

We use this NSGPR to achieve a more accurate and stochastic RUL prediction [37]. In the next sections, we discuss the datasets used in this work, the computational experiments, and the results obtained with VisPro.

## 3. Bearing prognostic datasets

In order to validate VisPro and benchmark its performance against other RUL prediction methods, we use the bearing experimental dataset provided by the "PHM12" challenge [30], which was originally acquired through "PRONOSTIA" experiment platform developed by FEMTO-ST Institute. Figure 3 illustrates the setup of the experiment platform,

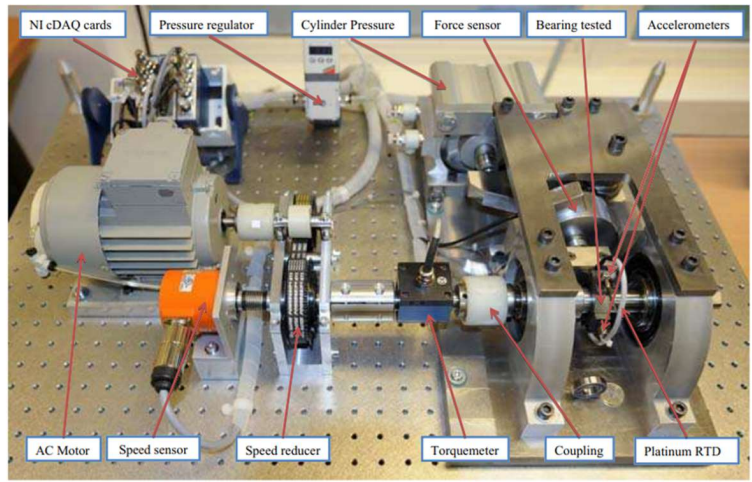

Figure 3. PRONOSTIA experimental platform setup [30]

The platform is self-contained with three principal components, including the rotating, the loading, and the measurement parts. For data acquisition, two types of sensors were used, accelerometers and resistance temperature detectors (RTD). The accelerometer and temperature measurements are sampled at frequencies of 25.6 kHz and 10 Hz respectively. Given the slow nature of temperature variations and the fact that the higher frequency vibration data can immediately reflect the health state of the bearing, it was decided to use the accelerometer measurements in our experiment to validate our prognostic model [14-20].

The data collection was conducted for a duration of 0.1s every 10s interval. Table 1 shows the three different working conditions of the experiment. A total of 17 bearings have been examined under three different operating conditions. In the first condition, seven bearings were examined under a load of 4000 N and 1800 rpm. In the second operating condition, another seven bearings were examined, under a load of 4200 N and 1650rpm. Finally, for the third condition, three bearings were examined under a load of 5000 N and 1500 rpm. In each condition, the data from two bearings are used as a training dataset, while others are the testing datasets. These are listed in Table 2. Under each operating condition, we train VisPro with the training subset, for example with Bearings 1_1 and 1_2 in condition 1, and we test the trained model with the corresponding testing subset, for example Bearings 1_3, 1_4, 1_5, 1_6, and 1_7 in condition 1.

Table 1. PHM12 dataset experimental conditions

| Working condition | Rotate speed (rpm) | Load (N) |
|---|---|---|
| 1 | 1800 | 4000 |
| 2 | 1650 | 4200 |
| 3 | 1500 | 5000 |

Table 2. Training and testing subsets of PHM12 prognostic dataset

| Dataset | Condition 1 | Condition 2 | Condition 3 |
|---|---|---|---|
| Training set | Bearing 1_1 | Bearing 2_1 | Bearing 3_1 |
|  | Bearing 1_2 | Bearing 2_2 | Bearing 3_2 |
| Testing set | Bearing 1_3 | Bearing 2_3 | Bearing 3_3 |
|  | Bearing 1_4 | Bearing 2_4 |  |

|  |  |
|---|---|
| Bearing 1_5 | Bearing 2_5 |
| Bearing 1_6 | Bearing 2_6 |
| Bearing 1_7 | Bearing 2_7 |

We use Bearing 1_2 from the training set to illustrate the details of the whole run-to-failure trajectories of the vibration signal and the actual RUL evolution with time. The evolution of the vibration signal along with its standard deviation (STD) showing the vibration magnitude [22] are provided in Fig. 4, and the ground truth of the RUL development is in Fig. 5.

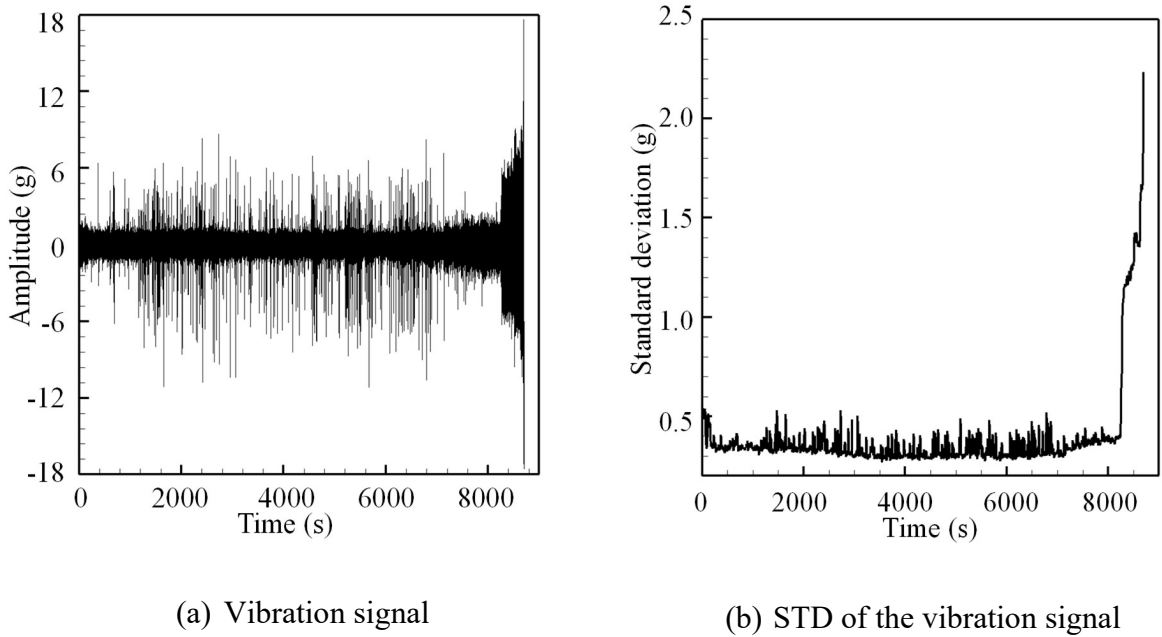

(a) Vibration signal        (b) STD of the vibration signal

Figure 4. Bearing 1_2 vibration and STD development in the training dataset

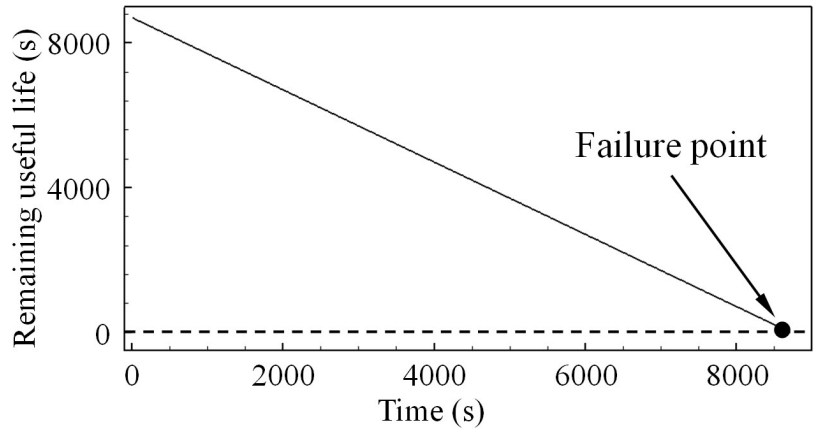

Figure 5. RUL as a function of time of Bearing 1_2 in the training dataset

Figure 4 illustrates the temporal evolution of the original vibration signal and its STD. After operating for about 8000s, the vibration and its STD increase sharply till the bearing fails. The RUL ground truth in Fig. 5 is only available for the bearings in the training datasets: knowing a bearing has failed at, say 8710s, when it reached 8000s, its RUL was 710s and shrunk linearly with further operation. If this bearing equipped a gas turbine or a jet engine for example, failure to detect that its RUL was shrinking dangerously may lead to a catastrophic outcome.

In Fig. 6, we illustrate the incomplete trajectories of the vibration signal of Bearing 1_6 in the testing dataset.

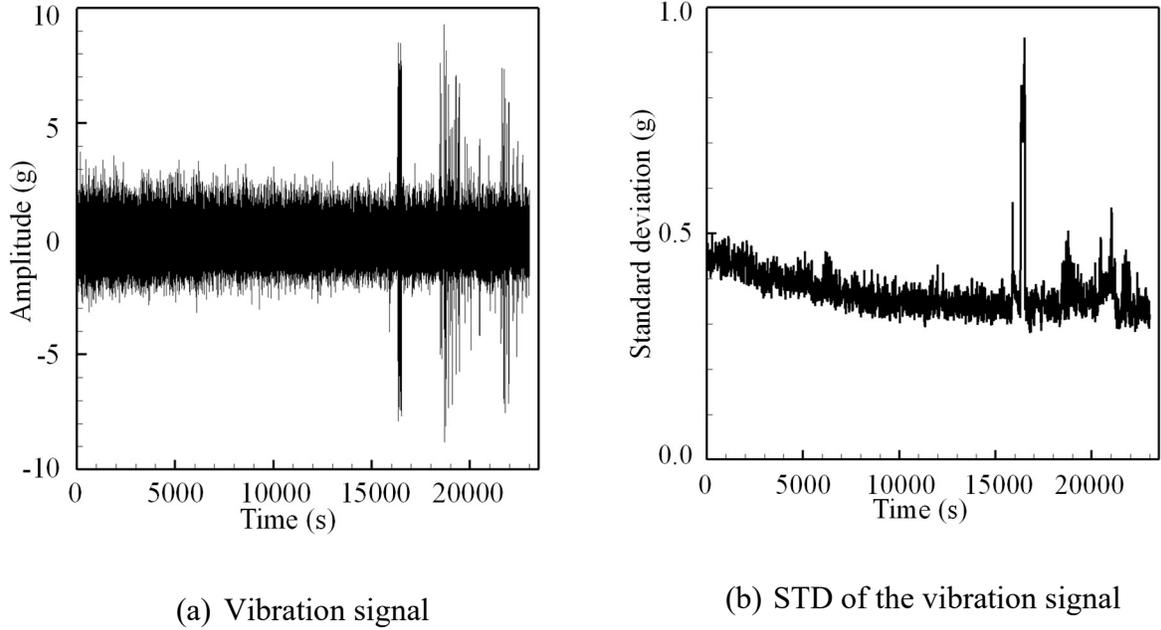

(a) Vibration signal  (b) STD of the vibration signal

Figure 6. Bearing 1_6 vibration and STD development in the testing dataset

By comparing the testing vibration signal in Fig. 6 with that of the training in Fig. 4, we note that the vibration signal in the testing dataset is truncated before the failure occurs.

Before getting to the results in the next section, we discuss the performance metric used to evaluate the RUL prediction accuracy of VisPro and other RUL prognostic models.

We use VisPro to predict the RUL of the testing Bearings, $\tilde{y}_i$ ($i \in \{1\_3, 1\_4, ..., 3\_3\}$). We then compare the estimated RUL at the time of truncation, $t_c$, with the ground truth $y_i$ ($i \in \{1\_3, 1\_4, ..., 3\_3\}$) provided by the testing dataset. To compare the RUL prediction between different models [14-20], we calculate a widely used residual index or prognostic score function to measure the RUL prediction accuracy. The percentage error of Bearing $i$ is calculated by Eq. 8, and the prognostic score function proposed by PHM12 [30] is provided in Eq. 9. A higher scoring function indicates smaller prediction error and more accurate model prediction.

$$Er_i = \frac{y_i - \hat{y}_i}{y_i} \times 100\% \tag{8}$$

$$A_i = \begin{cases} \exp\left(-\ln(0.5) \times \left(\frac{Er_i}{5}\right)\right) & if\ Er_i \leq 0 \\ \exp\left(+\ln(0.5) \times \left(\frac{Er_i}{20}\right)\right) & if\ Er_i > 0 \end{cases} \tag{9}$$

$$Score = \frac{1}{11}\sum_{i=1}^{11} A_i$$

## 4. Results and discussion

In this section, we examine the computational results obtained with VisPro using the PHM12 dataset. We first provide the results for the TFA and examine the temporal evolution of the bearing degradation in the time-frequency 2D space. Next, we provide a single bearing example in the testing dataset to illustrate the process and outcome obtained with model. Finally, we provide the RUL predictions for the entire testing dataset, and we compare the prediction accuracy of our model with that of other best-in-class models [14-20].

### 4.1. Results for time-frequency analysis

We use Bearing 1_2 as an example to demonstrate the vibrational signal evolution in the time-frequency domain over the entire life period of this training bearing. The original time streaming vibrational signal and its STD are shown in Fig. 4, and its evolution in the time-frequency domain is shown in Fig. 7 at four illustrative times, $t = 10s, 2910s, 5810s, 8710s$.

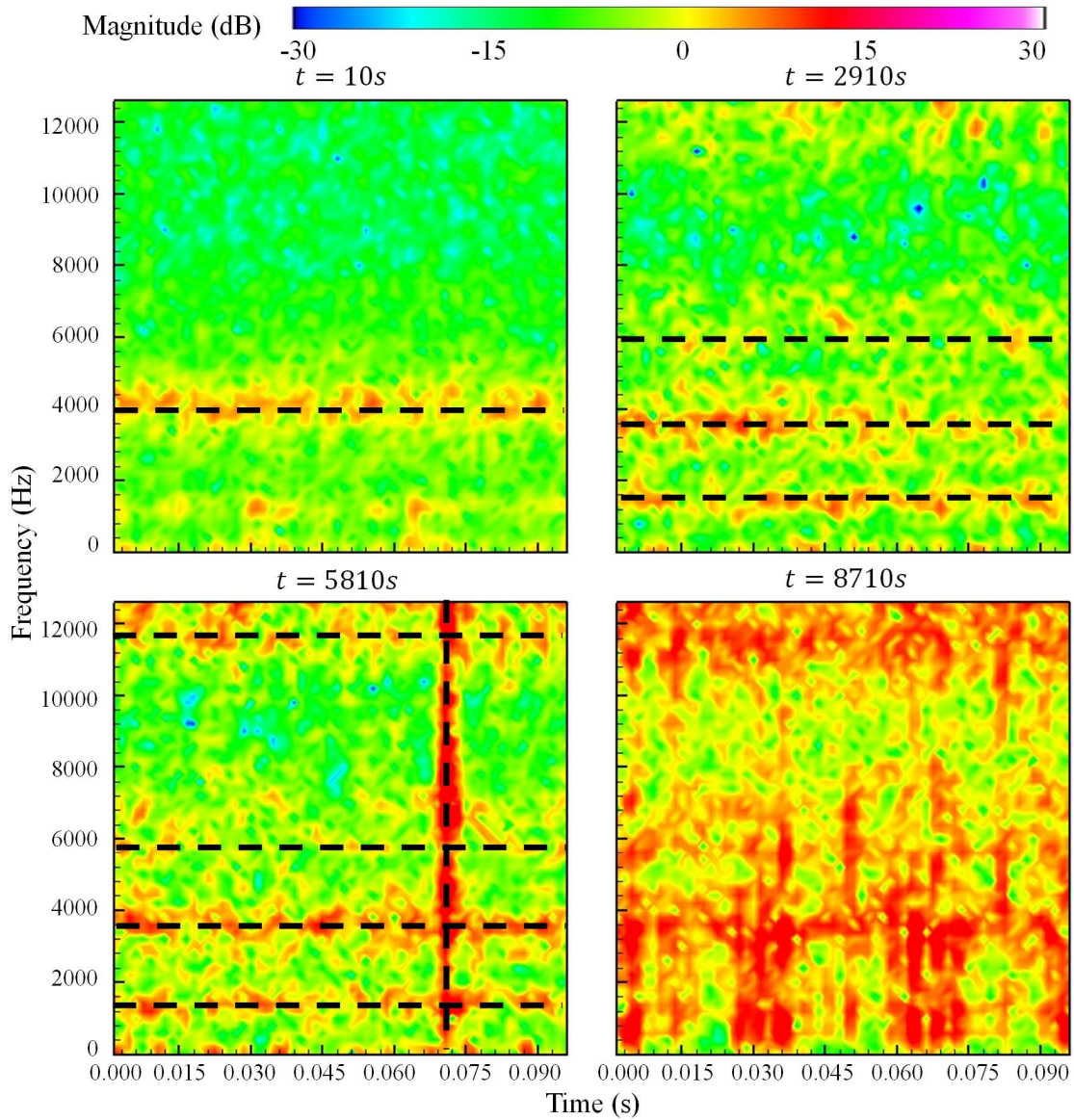

Figure 7. Bearing 1_2 vibration signal evolution in the time-frequency domain

The TFA displayed in Fig. 7 teases out more information that is not readily visible in the time domain (Fig. 4). For example, Bearing 1_2 clearly displays different phases in its degradation as seen in this four-panel figure. In the top-left panel, the vibration signal in the time-frequency domain displays a persistent 4000Hz signal (labeled by the dashed line). Although a snapshot is shown at $t = 10s$, this pattern persists until time $t = 1010s$. It likely reflects a nominal condition of the bearing. In the top-right panel, we observe a bifurcation in the time-frequency domain: three characteristic frequencies around $1800Hz$, $3800Hz$, and $6000Hz$ appear and persist in time. Although a snapshot is shown at $t = 2910s$, this pattern persists from time $t = 1010s$ until $t = 5000s$. This is likely indicative of the onset of some degradation. In the bottom-left panel, two new features appear in the time-frequency domain: a new persistent high frequency at $11800Hz$, and a broadband impulsive event [22, 41] at $t = 5810.07s$. This is the vertical dashed line in the bottom-left panel, and it reflects the occurrence of a very short event with frequencies across the entire range captured by the sensors in PRONOSTIA. This is likely an indication of a more serious degradation event. Finally, in the bottom-right panel, the vibration magnitude becomes significant over the entire frequency range and it indicates that the failure of the bearing is imminent. VisPro will "learn" from these patterns in the time-frequency domain and quantify their association with the time to failure in its RUL prediction.

Figure 7 and the previous observations were for a training bearing for which the entirety of the vibration signals are available up to the time to failure. The situation with the testing bearing is different. For example, with Bearing 1_6, the available vibration signal evolution and its STD development are shown in Fig. 6, and its TFA result at truncation time ($t = 230200s$) is shown in Fig. 8.

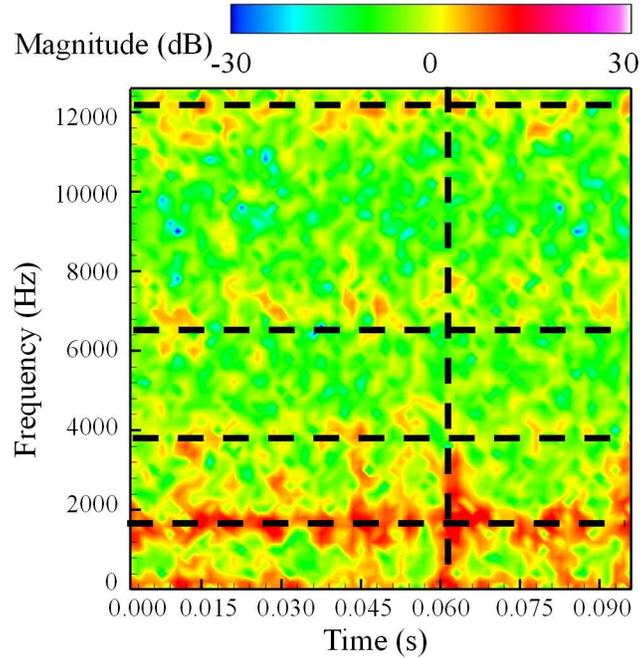

Figure 8. Bearing 1_6 vibration signal development in time-frequency domain at truncation time

As shown in Fig. 8, the testing signal in the time-frequency domain displays the following features: (1) persistent frequency characteristics line at $12000Hz$, $6000Hz$, $3800Hz$, and $1800Hz$; and (2) a (mild) temporal (vertical) characteristic line around $t = 230200 + 0.06$s. We note that Fig. 8 shows that the high frequency line at $12000Hz$ and the temporal line are not as fully developed and clearly delineated as in the bottom-left panel in Fig. 7.

### 4.2. RUL prediction results for Bearing 1_6 in the testing dataset

The incomplete vibration signal trajectory and the STD evolution of this Bearing were provided in Fig. 6, and the time-frequency analysis at the truncation time $t = 23020s$ in Fig. 8. No run-to-failure data for this bearing is available. The VisPro model predicts the mean RUL shown in Fig. 9 along with the uncertainty associated with it in the form of a 90% confidence interval. The 80% and 95% confidence intervals are provided in Appendix D. Their estimations are similar and are not included here to avoid visual clutter.

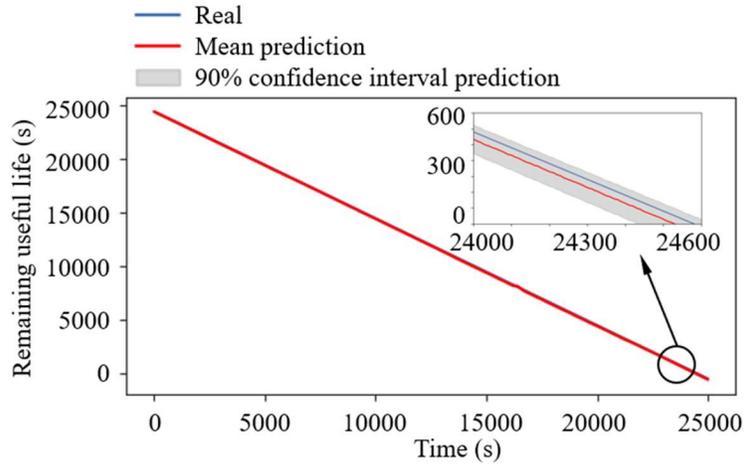

Figure 9. The RUL estimation of the VisPro model of the testing Bearing 1_6

At the time when the feature data is no longer available (truncation time $t_c = 23020s$), VisPro estimates the bearing to have a RUL $\tilde{y}_{1\_6} = 1400s$. This is the model's estimated remaining time before the bearing fails. In other words, failure is expected at time $t = 23020 + 1400 = 24420s$. The ground truth for this bearing's RUL at truncation time is provided by PRONOSTIA, and it is $y_{1\_6} = 1460s$. True failure, therefore, occurs at time $t = 24480s$. This difference is used to calculate the percent error (Eq. 9) and to benchmark the performance and accuracy of VisPro against other RUL predictive models. We note that this prediction is markedly accurate, and the 90% confidence interval contains the ground truth, as shown in Fig. 9. The lower bound and upper bound predictions are 1310s and 1500s, respectively and, the uncertainty bound includes the ground truth value.

## 4.3. RUL predictions for the entire testing dataset

We provide in Fig. 10 the RUL predictions for all the bearings in the testing dataset under different working conditions. The figure also includes the confidence intervals and the ground truth for each bearing. The results for all testing bearings are listed and provided in Table 3.

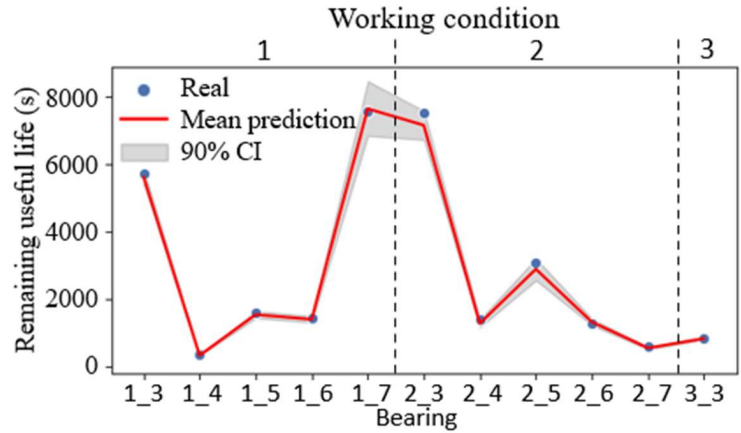

Figure 10. RUL predictions for the entire testing Bearings of different working conditions

Table 3. RUL predictions, ground truth values, and truncation time for entire testing Bearings

| Testing Bearing | RUL prediction at truncation time (s), $\tilde{y}_i$ | Ground truth RUL at truncation time (s), $y_i$ | 90% confidence interval, [lb; ub] | Truncation time (s), $t_c$ |
|---|---|---|---|---|
| 1_3 | 5630 | 5730 | [5420; 5860] | 18010 |
| 1_4 | 330 | 339 | [320; 340] | 11380 |
| 1_5 | 1540 | 1610 | [1440; 1650] | 23010 |
| 1_6 | 1400 | 1460 | [1310; 1500] | 23010 |
| 1_7 | 7650 | 7570 | [6850; 8470] | 15010 |
| 2_3 | 7160 | 7530 | [6730; 7590] | 12010 |
| 2_4 | 1280 | 1390 | [1170; 1400] | 6110 |
| 2_5 | 2890 | 3090 | [2560; 3210] | 20010 |
| 2_6 | 1320 | 1290 | [1220; 1420] | 5710 |
| 2_7 | 550 | 580 | [510; 590] | 1710 |
| 3_3 | 830 | 820 | [800; 870] | 3510 |

Figure 10 indicates that the RUL predictions are accurate and within the vicinity of the ground truth values, all the latter are within the 90% confidence intervals. This suggests that the uncertainty quantification is valid and informative. The average size of the 90% confidence interval for the RUL prediction with VisPro using the PROGNOSTIA testing dataset is 415s. All these intervals overlap with the ground truth. This can provide one useful benchmark for other researchers to gauge and compare their uncertainty quantification models for bearing RUL predictions. The prognostic score function for all the testing bearing dataset is:

$$Score = \frac{1}{11}\sum_{i=1}^{11} A_i = 0.84$$

How good or bad is this performance of the VisPro? To answer this question, it is useful to provide context and examine the performances of other best-in-class data-driven predictive models from recent literature. These models are trained on the same training dataset and tested on the same testing dataset. The results are provided in Table 4.

Table 4. Comparison of RUL prediction RMSE value of VisPro and other data-driven models

| Bearing | Current time | Actual RUL | Er% | | | | | | | |
|---|---|---|---|---|---|---|---|---|---|---|
| | | | Sutrisno (2012) [14] | Hong (2014) [15] | Lei (2016) [16] | Guo (2017) [17] | Yoo (2018) [18] | Wang (2018) [19] | Zhang (2021) [20] | VisPro |
| 1_3 | 18010 | 5730 | 37 | -1.04 | -0.35 | 43.28 | 1.05 | -1.05 | 2.27 | 1.75 |
| 1_4 | 11380 | 339 | 80 | -20.94 | 5.6 | 67.55 | 20.35 | -17.99 | 5.6 | 2.94 |
| 1_5 | 23010 | 1610 | 9 | -278.26 | 100 | -22.98 | 11.48 | 21.74 | 12.42 | 4.35 |
| 1_6 | 23010 | 1460 | -5 | 19.18 | 28.08 | 21.23 | 34.93 | 6.16 | 10.96 | 4.11 |
| 1_7 | 15010 | 7570 | -2 | -7.31 | -19.55 | 17.83 | 29.19 | 7.79 | -22.46 | -1.06 |
| 2_3 | 12010 | 7530 | 64 | 10.49 | -20.19 | 37.84 | 57.24 | 43.03 | 0.99 | 4.91 |
| 2_4 | 6110 | 1390 | 10 | 51.8 | 8.53 | -19.42 | -1.44 | 1.44 | 5.76 | 7.91 |
| 2_5 | 20010 | 3090 | -440 | 28.8 | 23.3 | 54.37 | -0.65 | 18.77 | 25.89 | 6.47 |
| 2_6 | 5710 | 1290 | 49 | -20.93 | 58.91 | -13.95 | -42.64 | 2.33 | -10.85 | -2.33 |
| 2_7 | 1710 | 580 | -317 | 44.83 | 5.17 | -55.17 | 8.62 | -3.45 | 1.72 | 5.17 |
| 3_3 | 3510 | 820 | 90 | -3.66 | 40.24 | 3.66 | -1.22 | 13.41 | -3.66 | -1.22 |
| Mean | N/A | N/A | 100.27 | 44.28 | 28.18 | 32.48 | 18.96 | 12.47 | 9.32 | **3.00** |
| STD | N/A | N/A | 173.28 | 90.29 | 35.41 | 37.57 | 23.59 | 15.9 | 12.57 | **3.18** |
| Score | N/A | N/A | 0.31 | 0.36 | 0.43 | 0.26 | 0.57 | 0.62 | 0.64 | **0.84** |

Table 4 indicates that the VisPro outperforms other data-driven predictive models based on the mean and STD of percent error and the prognostic score function:

- VisPro is over three times more accurate than the second-best model [20], the mean percent error for the latter is 3.00 and 9.32 for the latter. The performance differential with other models varies from 3 to 30 with an average about an order of magnitude across these alternate models.

- The STD of VisPro based on this testing dataset is four times smaller than the second-best model 3.18 versus 12.57 respectively. This smaller STD value indicates that VisPro is more robust to different working conditions compared with alternate models[1].

- The prognostic score function of VisPro is 0.84 about a third higher than the second-best model.

---

[1] It can be seen in Table 4 that for the bearing 2_3, the prediction accuracy of VisPro is worse than Ref. [20]. This is likely due to the random nature of the prediction performance in that reference with a high STD. For example, Ref. [20] has small error for bearing 2_3, but a significantly large error for bearing 2_5. VisPro achieves a more consistent RUL prediction accuracy with small mean and STD of prediction error for all testing bearings.

Overall, these results indicate that VisPro robustly outperforms competing RUL prediction models. The reasons for this excellent performance are threefold. First, the use of TFA preprocessing of the time streaming vibration signal teases out more hidden information from the original signal in the time-frequency 2D domain. Second, the Pro-SQN DL prognostic model is complex enough to capture the effect of the input TFA layer on the RUL estimation. Third, the NSGPR with dot product and local length scale kernels delivers on the advantages of learning the dynamics of the incomplete *RUL* predictions in the testing dataset, and it ultimately provides a more accurate and robust RUL predictions with uncertainty quantification.

## 5. Conclusion

In this work, we addressed some of the current challenges in data-driven RUL prediction of rotating machinery through vibration signal by capturing the non-stationarity of system degradation and RUL prediction and extracting more useful information from the input data. We devised a highly accurate RUL prediction model with uncertainty quantification, termed VisPro, which integrates and leverages the advantages of time-frequency analysis, deep learning SqueezeNet, and nonstationary Gaussian process regression. We examined and benchmarked our model against other advanced data-driven RUL prediction models using PHM12 Bearing vibration dataset. Our computational experiments showed that: (1) the VisPro predictions are highly accurate and provide significant improvements over competing prediction models; (2) the RUL uncertainty bounds are valid and informative.

This work should be considered in light of its limitations, and these constitute fruitful venues for future work. First, our RUL prediction model was only validated using the PHM12 Bearing prognostic dataset; more extensive testing is needed and will be conducted in the future to further challenge or confirm its advantages in other applications. Second, other (unsteady) kernels for the NSGPR than the one used here can be examined in the future and their contributions to the overall model performance assessed.

**Appendix A. Examining the effectiveness of NSGPR**

In this appendix, we examine the effectiveness of the NSGPR in our RUL prediction model by comparing the RUL prediction with and without NSGPR. The RUL predictions without NSGPR are taken from the last step prediction of the Pro-SQM network in the testing dataset. The RUL prediction without NSGPR the testing Bearing 1_2 is shown in Fig. A. 1.

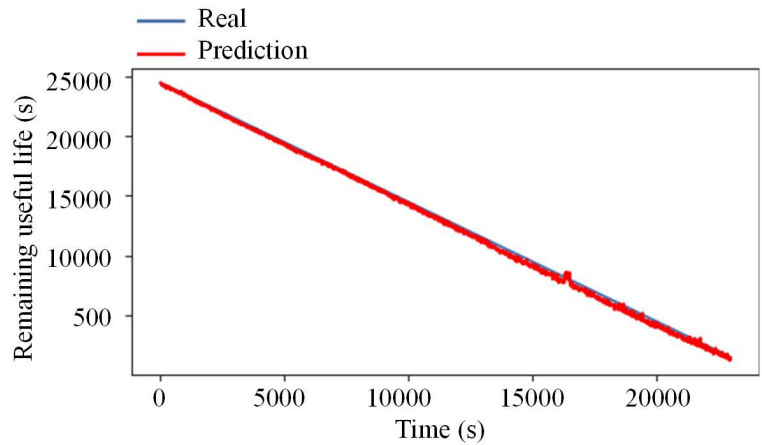

Figure A. 1. The RUL estimation of the DL network of the testing Bearing 1_2

First, we note that the RUL estimation curve is oscillating more significantly compared with the RUL prediction with NSGPR as shown in Fig. 9. Second, the RUL prediction is 1360s, which has a larger error than that of the prediction with NSGPR. The RUL predictions without NSGPR of overall testing Bearings are shown in Fig. A.2.

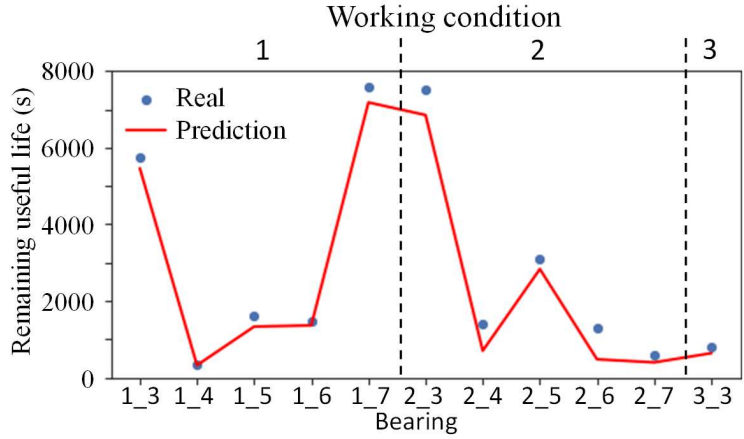

Figure A. 2. The RUL estimation of all Bearings in the testing set of DL network

First, the RUL prediction in Fig. A. 2, does not provide the lower and upper bounds, and the use of NSGPR supports our model with uncertainty quantification. Second, in this work, we use the scoring function as calculated in Eq. 9 to measure the accuracy of the model prediction. The scoring function for RUL predictions without NSGPR is 0.59. The scoring function of VisPro predictions is 0.84, which is 42% higher than that of the predictions without NSGPR.

## Appendix B. Structure and weight parameters of the Pro-SQN

Here, we introduce the details of the output size, memory requirement per image, and the number of the weight of Pro-SQN as shown in Table B.1. We use 32-bit floating numbers for variables in the data processing that one variable takes $\frac{32}{8} = 4$ bytes. According to Table B.1, the number of weights is 1,187M and it possesses 0.594Mb on memory for the model itself. The memory usage is the hard requirement of the hardware and weight size indicates the training requirement of the model.

Table B.1. The output size, hardware memory, and weight size of the fire detection SqueezeNet discriminator

| Layer | Output size | Memory/image | Weights number |
|---|---|---|---|
| Input | $[64 \times 64 \times 1]$ | $64 \times 64 \times 1 \times 4 = 16{,}384$ | N/A |
| Conv1 | $[30 \times 30 \times 32]$ | $30 \times 30 \times 32 \times 4 = 115{,}200$ | $(6 \times 6 \times 1) \times 32 = 1{,}152$ |
| Pool | $[15 \times 15 \times 32]$ | $15 \times 15 \times 32 \times 4 = 28{,}800$ | 0 |
| Fire2 | $[15 \times 15 \times 64]$ | $15 \times 15 \times (16 + 32 + 32) \times 4 = 72{,}000$ | $(1 \times 1 \times 32) \times 16 + (1 \times 1 \times 16) \times 32 + (3 \times 3 \times 16) \times 32 = 5{,}632$ |
| Fire3 | $[15 \times 15 \times 128]$ | $15 \times 15 \times (16 + 64 + 64) \times 4 = 129{,}600$ | $(1 \times 1 \times 64) \times 16 + (1 \times 1 \times 16) \times 64 + (3 \times 3 \times 16) \times 64 = 11{,}264$ |
| Pool | $[7 \times 7 \times 128]$ | $7 \times 7 \times 128 \times 4 = 25{,}088$ | 0 |
| Fire4 | $[7 \times 7 \times 256]$ | $7 \times 7 \times (32 + 128 + 128) \times 4 = 56{,}448$ | $(1 \times 1 \times 128) \times 32 + (1 \times 1 \times 32) \times 128 + (3 \times 3 \times 32) \times 128 = 45{,}056$ |
| Fire5 | $[7 \times 7 \times 256]$ | $7 \times 7 \times (32 + 128 + 128) \times 4 = 56{,}448$ | $(1 \times 1 \times 256) \times 32 + (1 \times 1 \times 32) \times 128 + (3 \times 3 \times 32) \times 128 = 49{,}152$ |
| Pool | $[3 \times 3 \times 256]$ | $3 \times 3 \times 256 \times 4 = 9{,}216$ | 0 |
| Fire6 | $[3 \times 3 \times 384]$ | $3 \times 3 \times (48 + 192 + 192) \times 4 = 15{,}552$ | $(1 \times 1 \times 256) \times 48 + (1 \times 1 \times 48) \times 192 + (3 \times 3 \times 48) \times 192 = 104{,}448$ |
| Fire7 | $[3 \times 3 \times 384]$ | $3 \times 3 \times (48 + 192 + 192) \times 4 = 15{,}552$ | $(1 \times 1 \times 384) \times 48 + (1 \times 1 \times 48) \times 192 + (3 \times 3 \times 48) \times 192 = 110{,}592$ |
| Fire8 | $[3 \times 3 \times 512]$ | $3 \times 3 \times (64 + 256 + 256) \times 4 = 20{,}736$ | $(1 \times 1 \times 384) \times 64 + (1 \times 1 \times 64) \times 256 + (3 \times 3 \times 64) \times 256 = 188{,}416$ |
| Fire9 | $[3 \times 3 \times 512]$ | $3 \times 3 \times (64 + 256 + 256) \times 4 = 20{,}736$ | $(1 \times 1 \times 512) \times 64 + (1 \times 1 \times 64) \times 256 + (3 \times 3 \times 64) \times 256 = 196{,}608$ |
| Conv10 | $[3 \times 3 \times 1024]$ | $3 \times 3 \times 1024 \times 4 = 36{,}864$ | $(1 \times 1 \times 512) \times 1024 = 524{,}288$ |
| Pool | $[1 \times 1 \times 1024]$ | $1 \times 1 \times 1024 \times 4 = 4{,}096$ | 0 |
| Den1 | $[1 \times 1 \times 4]$ | $1 \times 1 \times 4 \times 4 = 16$ | $1024 \times 4 = 4{,}096$ |
| Den2 | $[1 \times 1 \times 100]$ | $1 \times 1 \times 100 \times 4 = 400$ | $5 \times 100 = 500$ |
| Den3 | $[1 \times 1 \times 30]$ | $1 \times 1 \times 30 \times 4 = 120$ | $100 \times 30 = 3{,}000$ |
| Den4 | $[1 \times 1 \times 1]$ | $1 \times 1 \times 1 \times 4 = 4$ | $30 \times 1 = 30$ |
| Overall | N/A | $623{,}260 B = 0.594 Mb$ | $1{,}244{,}234 = 1.187 M$ ($4.748 Mb$ on memory) |

Then, we introduced the fire model as shown in Fig. B. 1 since it is extensively used in the SqueezeNet, where $s_{1x1}$, $e_{1x1}$, and $e_{3x3}$ stand for the number of squeeze layers, the number of $1 \times 1$ expand layer, and the number of $3 \times 3$ expand layers, respectively.

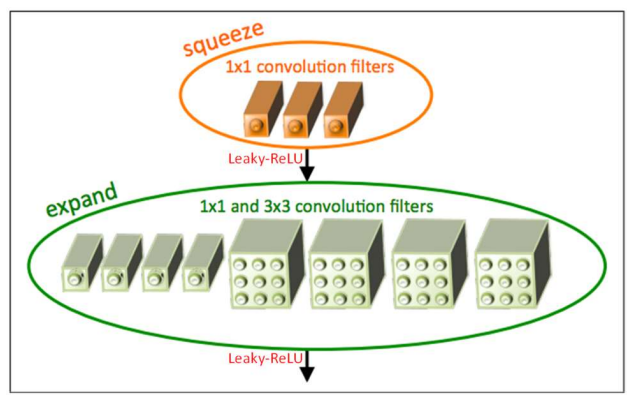

Figure B.1. Organization of convolution filters in the fire model. In this example, $\boldsymbol{s_{1x1}} = \boldsymbol{3}$, $\boldsymbol{e_{1x1}} = \boldsymbol{4}$, and $\boldsymbol{e_{3x3}} = \boldsymbol{4}$. We illustrate the convolution filters but not the activations [33].

In our fire detection SqueezeNet, we set $s_{1x1}$, $e_{1x1}$, and $e_{3x3}$ as 1. We switch the activation function of the fire model from ReLU in the original model to LeakyReLU for more nonlinearity and preventing vanishing gradient problem for negative input.

**Appendix C. Examining the effectiveness of the local length scale kernel**

In this appendix, we examine the effectiveness of the local length scale kernel in our RUL prediction model. In order to demonstrate the effectiveness of the local length scale, we compare our results with a prediction with dot product and squared exponential kernel, which has a universal length scale. The RUL predictions without local length scale kernel of overall testing Bearings are shown in Fig. C.1 and Table C. 1.

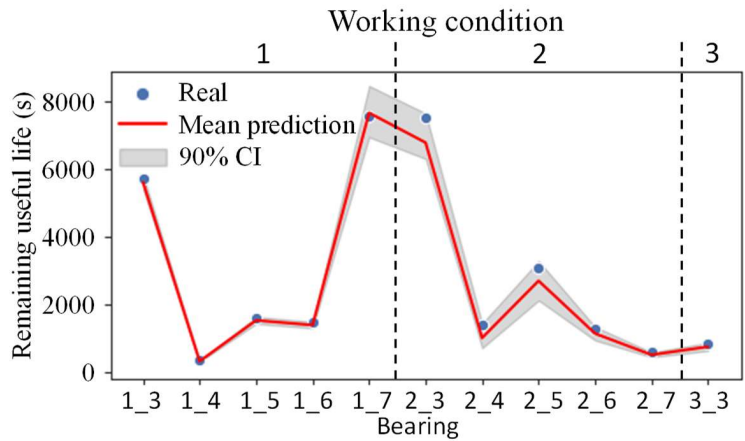

Figure C. 1. The RUL estimation of the entire Bearings in the testing set without local length scale kernel

Table C.1. Comparison of RUL prediction of VisPro with and without local length scale kernel

| Bearing | Current time | Actual RUL | Er% With local length scale | Er% Without local length scale |
|---|---|---|---|---|
| 1_3 | 18010 | 5730 | 1.75 | 2.09 |
| 1_4 | 11380 | 339 | 2.94 | 5.88 |
| 1_5 | 23010 | 1610 | 4.35 | 4.97 |
| 1_6 | 23010 | 1460 | 4.11 | 4.79 |
| 1_7 | 15010 | 7570 | -1.06 | -1.32 |
| 2_3 | 12010 | 7530 | 4.91 | 9.83 |
| 2_4 | 6110 | 1390 | 7.91 | 27.34 |
| 2_5 | 20010 | 3090 | 6.47 | 12.62 |
| 2_6 | 5710 | 1290 | -2.33 | 11.63 |
| 2_7 | 1710 | 580 | 5.17 | 12.07 |
| 3_3 | 3510 | 820 | -1.22 | 8.54 |
| Mean | N/A | N/A | **3.00** | 8.94 |
| STD | N/A | N/A | **3.18** | 7.10 |
| Score | N/A | N/A | **0.84** | 0.74 |

First, in Table C. 1, the RUL prediction with local length scale kernel has a significant advantage over the prediction with squared exponential kernel. The mean of the prediction error, STD of the error, and score function are improved by 66%, 55%, and 14%, respectively. Second, in Table C. 1, the RUL prediction error of Bearing 2_4 is significantly improved from 27.34 to 7.91 by using local length scale kernel. In order to investigate the details of this improvement, the RUL predictions for the testing Bearing 2_4 with and without local length scale kernel are compared in Fig. C. 2.

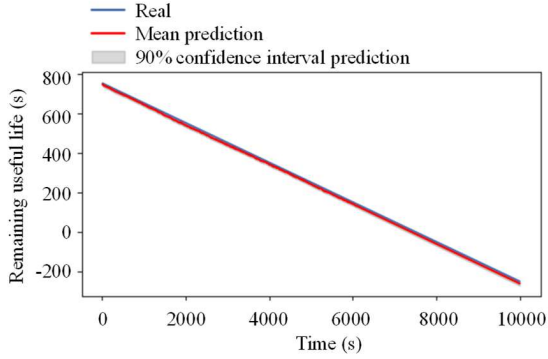 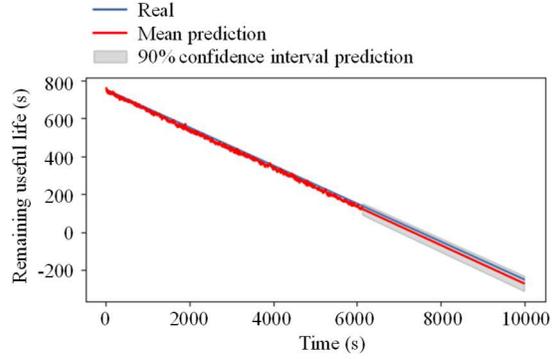

(a)             (b)

Figure C. 2. The RUL estimation of Bearing 2_4 (a) with local length scale kernel; (b) without local length scale kernel

Comparing the results of RUL prediction of Bearing 2_4 with local length scale kernel and without local length scale kernel (with squared exponential kernel), first, the prediction of without local length scale kernel is oscillating. Since the local length scale kernel considers the local smoothness, its prediction is less oscillating and more robust. This consequently improves the prediction accuracy of the B 2_4 in the testing dataset. Second, in Fig. C. 2, the uncertainty bound is larger after truncation time (6110s) for the prediction without local length scale kernel compared with that of the prediction with local length scale kernel. Consequently, the use of local length scale kernel in the NSGPR step provides a more precise RUL prediction with a tighter uncertainty bound.

**Appendix D. Uncertainty quantification of 80%, 90%, and 95% confidence interval**

In this appendix, we examine the uncertainty quantification results with 80%, 90%, and 95% confidence intervals. The prediction and uncertainty quantification results are shown in Fig. D. 1.

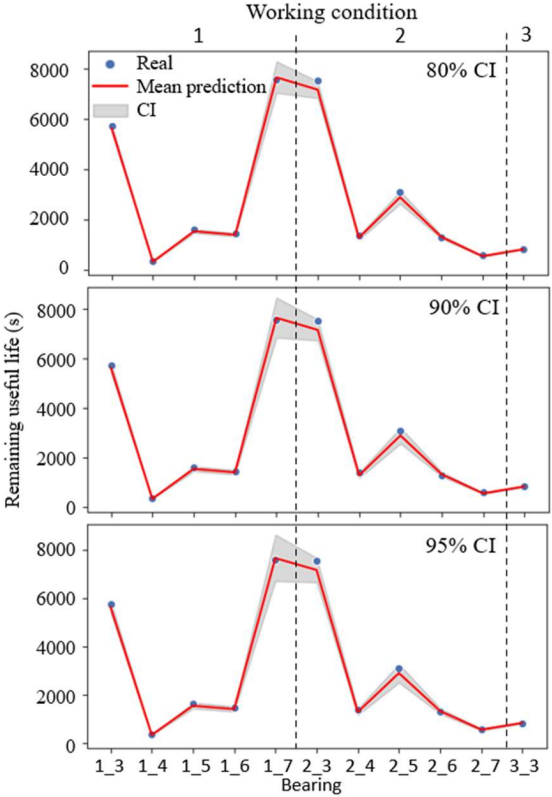

Figure D. 1. The RUL estimation of the entire Bearings in the testing set with 80%, 90%, and 95% confidence interval

Figure D. 1 shows that first the mean estimation of RUL is identical and does not vary with different confidence intervals. Second, virtually, the size of uncertainty shrinks with the decrease of its percentage. For example, 80% confidence interval is tighter compared with that of 90% and 95%. However, the uncertainty quantification has potential invalid cases that ground truths are outside the uncertainty bound for the 80% confidence interval. We calculate and summarize the average confidence interval size and invalid case number among all testing bearings for 80%, 90%, and 95% confidence intervals in Table D.1.

Table D. 1. The uncertainty quantification performance with 80%, 90%, and 95% confidence interval

| Confidence interval | Average confidence interval size | Invalid |
| --- | --- | --- |
| 80% | 322s | 3 |
| 90% | 415s | 0 |
| 95% | 494s | 0 |

Table D. 1 shows 80% confidence interval has the tightest uncertainty bound with the average size of 322s. However, it has 3 invalid cases that the real RULs are out of the estimated uncertainty bounds, and their uncertainty quantifications are not informative. Second, 90% has a tighter uncertainty bound compared with that of 95% without invalid case of uncertainty quantification. In this way, with the considerations of both average size and invalid case of the uncertainty quantification, 90% confidence interval is the optimal selection for VisPro. We use 90% confidence interval in Section 4 to discuss the RUL prediction and uncertainty quantification results.

## Acknowledgments

This work was supported in part by a Space Technology Research Institute grant from NASA's Space Technology Research Grants Program.